# Photoluminescence Quenching in Single-layer MoS$_2$ via Oxygen Plasma Treatment

Narae Kang[1,2], Hari P. Paudel[1,2], Michael N. Leuenberger[1,2,3], Laurene Tetard[1,2,*], and Saiful I. Khondaker[1,2,4,*]

[1]Nanoscience Technology Center (NSTC), [2]Department of Physics, [3]College of Optics and Photonics (CREOL), [4]School of Electrical Engineering and Computer Science, University of Central Florida, 12424 Research Parkway, Suite 400, Orlando, Florida, 32826, USA

ABSTRACT

By creating defects via oxygen plasma treatment, we demonstrate optical properties variation of single-layer MoS$_2$. We found that, with increasing plasma exposure time, the photoluminescence (PL) evolves from very high intensity to complete quenching, accompanied by gradual reduction and broadening of MoS$_2$ Raman modes, indicative of distortion of the MoS$_2$ lattice after oxygen bombardment. X-ray photoelectron spectroscopy study shows the appearance of Mo$^{6+}$ peak, suggesting the creation of MoO$_3$ disordered regions in the MoS$_2$ flake. Finally, using band structure calculations, we demonstrate that the creation of MoO$_3$ disordered domains upon exposure to oxygen plasma leads to a direct to indirect bandgap transition in single-layer MoS$_2$, which explains the observed PL quenching.

INTRODUCTION

The ability to controllably tailor the properties of a material is a key factor in the development of many novel applications. In the case of bulk semiconductors, creating and manipulating defects constitutes an essential element in controlling the electrical, magnetic, and optical properties of the host material.[1] Although the role of defects is well understood in bulk semiconductors, it has received little attention in emerging two-dimensional (2D) layered semiconductors, preventing their full exploitation for tailored 2D nanoelectronic and photonic devices. Graphene and graphene oxide are examples of the impact that defects can have on 2D materials. Pristine graphene, which contains no intrinsic defect, is well known for its extraordinary high mobility, and is of great importance for high frequency device applications.[2,3] However, its inherent lack of bandgap and low absorption of solar photons greatly limit its use in electronic and photonic devices. On the other hand, its solution processed counterparts, graphene oxide and reduced graphene oxide, have a large amount of defects, which lead to formation of a bandgap and open the way to many other applications in photodetectors, sensors, catalysis, and solar cell.[4-8]

Recently, layered transition metal dichalcogenides (TMDs) have emerged as important materials for 2D device engineering.[9-11] Molybdenum disulfide (MoS$_2$), composed of weak van der Waals bonded S-Mo-S units, offers a large intrinsic bandgap that is strongly dependent on the number of layers, with an indirect bandgap (1.2 eV) in bulk MoS$_2$ transitioning to a direct bandgap (1.8 eV) in single layer (SL).[12, 13] Although significant efforts have been placed in understanding the properties and applications of pristine SL $MoS_2$,[9-22] altering the properties of SL $MoS_2$ via creating and manipulating defects through external control is still in its infancy.[23-29] Such defect engineering can modify the optical properties, such as enhancing[23] or quenching[24] of PL, as well as tailor the electrical properties of SL $MoS_2$ [27].

Here we demonstrate a new and simple technique to alter the optical properties of the SL $MoS_2$ by creating molybdenum trioxide ($MoO_3$) defects by oxygen plasma treatment. We found that the PL of SL $MoS_2$ sheets gradually decreases from a high response to a complete quenching with increasing plasma exposure time. Raman spectra acquired on $MoS_2$ sheets treated with different plasma exposure durations show that the in-plane $E^1_{2g}$ and out-of-plane $A_{1g}$ mode intensities of $MoS_2$ gradually decreases while their position shifts in time, suggesting significant lattice distortions. In addition, a new peak at 225 $cm^{-1}$ appears as soon as the $MoS_2$ was exposed to $O_2$ plasma. Furthermore, X-ray photoelectron spectroscopy (XPS) study shows the appearance of $Mo^{6+}$ peak suggesting that PL quenching is related to the creation of $MoO_3$ disordered regions in the treated $MoS_2$ flake. Finally, using band structure calculations, we confirmed the $MoS_2$ bandgap evolution from direct to indirect in presence of $MoO_3$ defects in the lattice, which results in PL quenching. Our results show a new defect inducing technique to engineer the bandgap and control the optical properties of atomically thin $MoS_2$.

**EXPERIMENTAL METHODS**

*Sample preparation:* The single-layer (SL) $MoS_2$ was mechanically exfoliated from commercially available bulk $MoS_2$ (SPI Supplies) using adhesive tape and deposited on highly doped Si substrate with a thermally grown 250 nm thick $SiO_2$. The plasma treatment was carried out using a Plasma Etch (PE-50) system operated with power 100 W at 50 kHz. For plasma exposure, the chamber pressure was set at 250 mTorr and a gas mixture of Oxygen (20%) and Argon (80%) flowed at a constant rate of 15 sccm. For time-dependent plasma exposure, the samples were exposed at 1 s iterations and the optical measurements were performed after each plasma exposure.

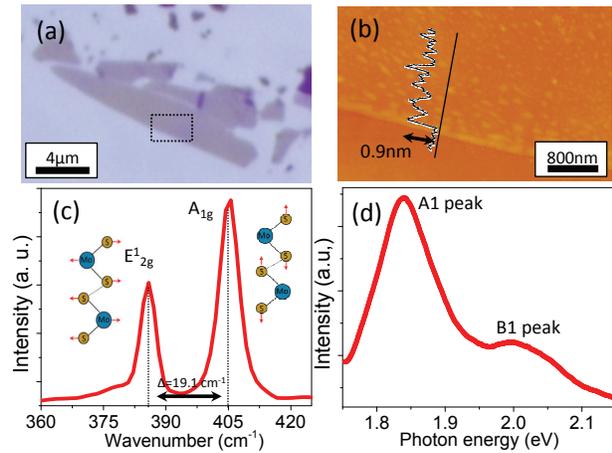

**Figure 1.** Single-layer (SL) $MoS_2$ characterization. (a) Optical, and (b) AFM image with height profile. (c) Raman spectrum, and (d) photoluminescence (PL) spectrum of exfoliated SL $MoS_2$ on $Si/SiO_2$ substrate.

*Characterization:* The PL measurements were carried out using a confocal Raman microscope (Witec alpha 300 RA) with a 100× objective and a Si detector (detection range up to ~2.2 eV). The Raman spectra measurements were performed on the same microscope using a 532 nm laser in ambient environment at room temperature. To ensure the reproducibility of the data, we followed a careful alignment and optimization protocol. In addition, the laser power was maintained below 1 mW to avoid any local heating and possible additional oxidation of the samples, independently of the integration time selected. The integration time was optimized to obtain a satisfactory signal-to-noise ratio.

*Theoretical calculation:* Density functional theory calculations (DFT) were performed to calculate the bandstructures for a SL of $MoS_2$, a SL of $MoO_3$, a SL of $MoS_2$ with O defects, and a SL of $MoS_2$ with $MoO_3$ defects. In each case we considered a mesh of 9×9×1 k-points in the Brillouin zone. The ion-electron interaction is described by the projected augmented wave (PAW) method and the exchange-correlation energy was calculated using the Perdew, Burke and Ernzerhof (PBE) approximation within the framework of the generalized gradient approximation (GGA). The grid point cutoff of 415 eV was used and a maximum force of 0.1 eV/Å on each atom was reached during the optimization process in all cases.

**RESULTS AND DISCUSSIONS**

The optical and atomic force microscopy (AFM) images of a SL $MoS_2$ flake is presented in Figure 1 (a) and (b), respectively. The flake has a height of 0.9 nm (Figure 1(b)) indicating a SL $MoS_2$. In addition, Raman spectrum (Figure 1(c)) shows two prominent peaks corresponding to the in-plane $E^1_{2g}$ and out-of-plane $A_{1g}$ vibrations of $MoS_2$ (Figure 1c inset), with a position difference (Δ) of 19.1 cm$^{-1}$, further confirming the SL thickness of the flake.[30, 31] Figure 1(d) shows the PL profile of the flake with a strong PL peak at 1.84 eV (A1) arising from the direct recombination of photo-generated electron-hole pairs, with the characteristic higher luminescence quantum efficiency of SL. The weaker peak at 2.02 eV (B1) corresponds to the energy split of the valence band spin-orbital coupling of $MoS_2$ occurring in presence of the $SiO_2$/Si substrate.[12]

Figure 2 shows the effect of increasing oxygen plasma exposure on the PL of SL $MoS_2$. The plasma exposure duration ranging from 1 to 6 s, with 1 s iterations, were labeled as t1 to t6. Figure 2(a) shows that the PL intensity decreases with increasing plasma exposure time. To accurately compare the PL intensities, all spectra were fitted with a Lorentzian function considering a constant baseline (details are presented in supplementary information Figure S1). Figure 2(b) shows the fitted curves where the curves representative of the A1 peak are shown in left panel and the curves representative of B1 peak are shown in right panel. Initially, the SL $MoS_2$ flake exhibited a A1 peak intensity of 30.50 CCD cts at 1.84 eV, which drastically weakened down to 8.91 CCD cts at t1, 4.87

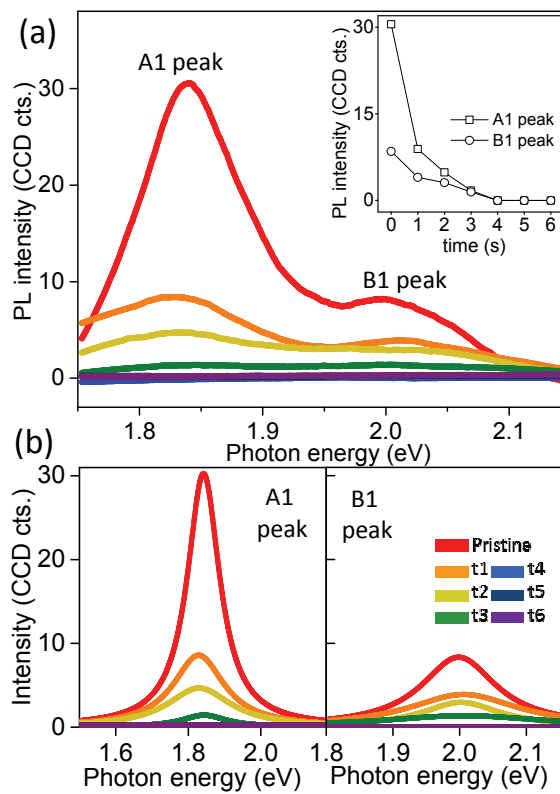

**Figure 2.** (a) Time-dependent photoluminescence (PL) spectra of plasma treated SL $MoS_2$. Inset shows the PL intensity of A1 and B1 peak with respect to the plasma exposure time from pristine SL $MoS_2$ (red curve) to t6 (purple curve). (b) Corresponding to A1 and B1 excitons, the peaks in PL spectra were fitted with Lorentzian functions.

CCD cts at t2 and 1.74 CCD cts at t3. After t3, the PL of A1 peak was fully quenched (Figure 2(a) inset). Similar behavior was observed for B1 peak (at 2.02 eV) with intensity at 8.51 CCD cts in the pristine flake, 4.02 CCD cts at t1, 3.09 CCD cts at t2 and 1.48 CCD cts at t3. The PL of peak B1 was also fully quenched after t3. Emitted radiations in PL are caused by radiative recombination upon photo-excitation. Radiative recombination is the most effective in direct bandgap semiconductors, such as SL $MoS_2$, as it only requires one step recombination for the electron to transition back to its equilibrium level. Therefore, the observed PL quenching suggests that the recombination process of pristine $MoS_2$ has been modified by the bombardment of oxygen, engendering lattice distortion and affecting the bandgap at 1.84 eV.

In a recent work involving electrical properties of $MoS_2$,[27] we showed that similar oxygen plasma treatment results in an exponential increase in resistance of SL $MoS_2$ by up to 4 orders of magnitude after 6 s treatment. XPS study carried out on the 6 s plasma treated sample showed peaks at 227 eV, 229.7 eV and 233.1 eV that have been attributed to binding energy of Mo $3d_{5/2}$, Mo $3d_{3/2}$, and S 2s electrons in Mo-S bond of the $MoS_2$ crystal (see supporting information Figure S2).[27, 32] An additional peak at 236.4 eV corresponding $Mo^{6+}$ state was observed, indicating the formation of $MoO_3$ defect region within the SL $MoS_2$ after plasma treatment.[27, 33] The formation of $MoO_3$ is also supported by the fact that, during oxygen plasma exposure, energetic O/Ar atoms hit the $MoS_2$ surface and the S atoms can move out of the lattice site, creating lattice vacancies. Due to the presence of oxygen in the plasma, oxidation takes place at the defect sites and the reaction can be described as $2MoS_2 + 7O_2 \rightarrow 2MoO_3 + 4SO_2$.[27] In order to explain the exponential increase of resistance with plasma exposure time, we hypothesized that the creation of $MoO_3$ in $MoS_2$ forms significant distortions of lattice and that such lattice distortions become more severe with increasing plasma exposure time. The same lattice distortion may also be responsible for the PL decay observed here.

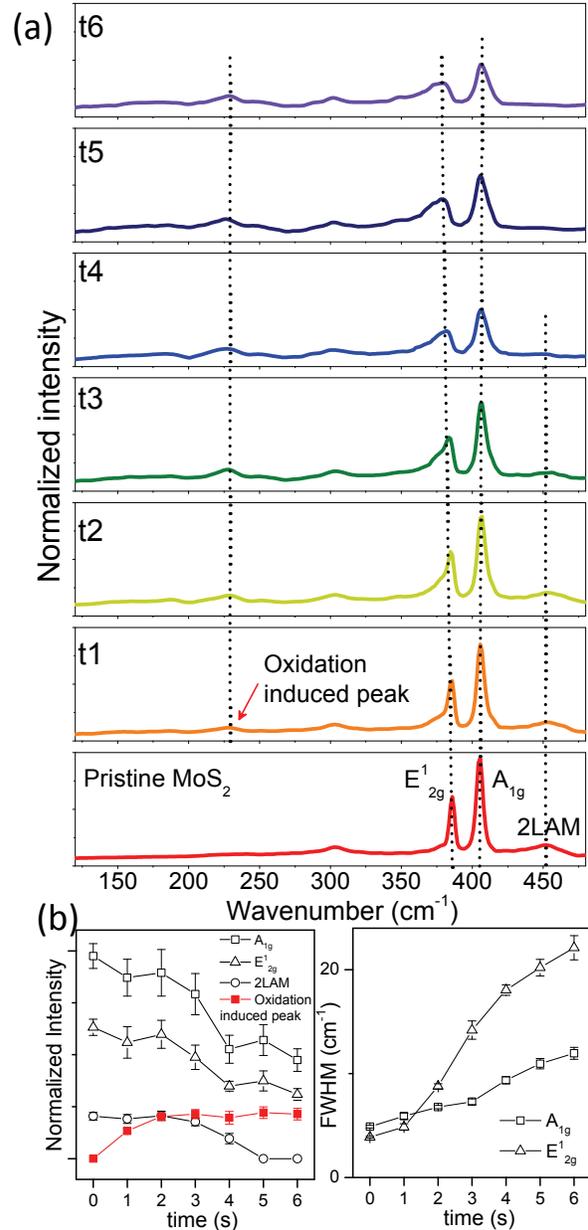

**Figure 3.** (a) Raman spectra of SL pristine $MoS_2$ and plasma-treated $MoS_2$ as function of plasma exposure time. For comparison, all the curves were plotted in the same y-scale (b) Evolution of normalized peak intensity, and FMHW (full-maximum half-width) for $A_{1g}$ (open square), $E^1_{2g}$ (open triangle), 2LAM (open circle), and oxidation induced (OI) peak (red closed square) peak as a function of plasma treatment time.

In order to confirm the lattice distortions caused by oxygen plasma and understand their formation in time, we systematically investigated localized Raman spectroscopy measurements of the $MoS_2$ flakes after incremental plasma exposure (pristine to t6). The results are presented in Figure 3 (a). The intensity of all Raman spectra was normalized over the amplitude of the Si peak at 520 cm$^{-1}$. At t1, the intensity of both $A_{1g}$ and $E_{2g}^1$ modes showed an 11% decrease compared to the intensity of pristine SL $MoS_2$ (red curve). In addition, from pristine SL $MoS_2$ to t1, position shifts were observed from 404.9 cm$^{-1}$ to 405.7 cm$^{-1}$ for the $A_{1g}$ mode and from 385.9 cm$^{-1}$ to 385.0 cm$^{-1}$ for the $E_{2g}^1$ mode. We also found width broadening from 4.9 cm$^{-1}$ to 5.91 cm$^{-1}$ for $A_{1g}$ and from 3.9 cm$^{-1}$ to 4.9 cm$^{-1}$ for $E_{2g}^1$ as a result of the treatment t1 on the pristine SL $MoS_2$ (Figure 3(b), right panel). The intensity of 2LAM decreased about 6% at t1 while its position shifted from 451.5 cm$^{-1}$ to 452.2 cm$^{-1}$. After t4, the intensity of $A_{1g}$ and $E_{2g}^1$ dropped down to 46% and 45% of the intensity values of pristine SL $MoS_2$, respectively. The corresponding positions were 406.8 and 380.5 cm$^{-1}$. The position shifts and the increase in width evolved until t6. The results are summarized in Figure 3(b). Overall, the position of $A_{1g}$ underwent a blue shift of up to 2.3 cm$^{-1}$, while the position of $E_{2g}^1$ red shifted up to 8.3 cm$^{-1}$ from their position in the pristine SL $MoS_2$ spectrum (Figure 3(a)). As a result, Δ was found varying from 19.1 cm$^{-1}$ in pristine $MoS_2$ to 29.7 cm$^{-1}$, which is larger than Δ measured in bulk $MoS_2$, signifying major changes in the material's properties and structure, including displacement of the Mo and S atoms from their original sites, due to the presence of O in the flake. As highlighted in the inset of Figure 1(c), shifts of $A_{1g}$, the out-of-plane optical vibration mode of S atoms, and $E_{2g}^1$, the in-plane optical vibration mode of the Mo-S bond, confirm the evolving lattice dynamics of the system. The continual decrease in intensity of all $MoS_2$ bands, together with the continual softening of the $A_{1g}$ phonon mode, and stiffening of $E_{2g}^1$ phonon mode, as a result of increasing exposure time, show the growing loss of $MoS_2$ crystal symmetry, and gradual increase of disorder in the lattice due to creation of oxygen defect sites,[33, 34] further confirmed by the increase in FWHM of $A_{1g}$ and $E_{2g}^1$ (Figure 3(b) right), and by the disappearance of the 2LAM mode. This provides strong evidence that PL quenching is induced by the significant lattice distortions in the treated flake.

Additionally, the oxidation induced (OI) peak at 225 cm$^{-1}$, which may indicate the formation of Mo-O bonds,[27] appeared as soon as the flake was bombarded with oxygen (at t1). According to XPS study (see supplementary material), the presence of $Mo^{6+}$ state in the treated flake suggests that Mo-O bonds lay in a flake of $MoS_2$ (which is still present as supported by the predominant $MoS_2$ peaks). However, the symmetry of the Mo-O bonds is expected to be different from that of bulk $MoO_3$, due to the in-plane interaction with Mo and S, which may explain the absence of the bulk $MoO_3$ Raman band in the spectra. Hence, the OI peak was not detected in pristine SL $MoS_2$ and appeared starting at t1 to finally increase to a level similar to that of 2LAM in pristine flakes (Figure 3(b)). No noticeable increase was found after t3, which could not be explained at this point, and should be the subject of further investigation. Nevertheless, the observations concur with the significant PL quenching after t3.

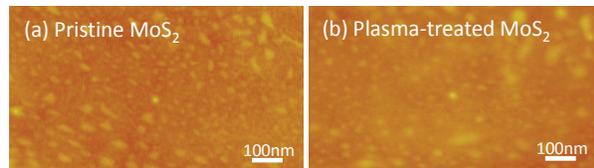

**Figure 4.** AFM images (a) before and (b) after plasma-treated SL $MoS_2$

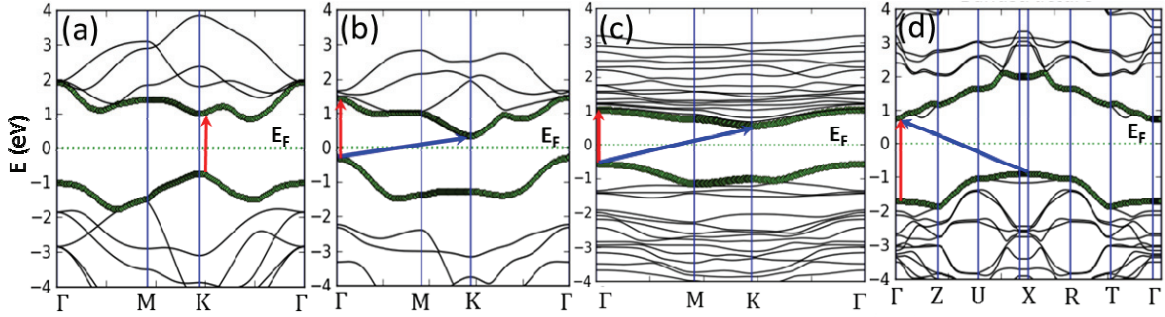

**Figure 5.** Band structure of (a) pristine SL MoS$_2$, (b) SL MoS$_2$ with one S atom replaced by one O atom, (c) SL MoS$_2$ with three S atoms replaced by three O atoms, and (d) a SL of MoO$_3$. SL of MoS$_2$ has direct bandgap at K point (a). The replacement of an S atom by an O atom changes the positions of the band extrema (b,c). The d-bands of MoS$_2$ at the band extrema are more hole type at the K point (a).

To ensure that the observed properties were not coming from any physical damage or etching of SL MoS$_2$, we acquired AFM images of the sample before and after 6s oxygen plasma treatment, as presented in Figure 4. As can be seen, the SL MoS$_2$ flake did not show any sign of etching or physical damage after plasma treatment.

Finally, we performed density functional theory (DFT) calculations to determine the bandstructure of pristine SL MoS$_2$, and MoS$_2$ after oxidation at the defect sites by considering SL MoS$_2$ with O defects and SL MoS$_2$ with MoO$_3$ defects along the high symmetry lines in the first Brillouin zone, as shown in Figure 5. Our band structure calculation presented in Figure 5 shows that for a pristine SL of MoS$_2$ doped with an O atom, the conduction band minimum (CBM) remains at the K point (as in pristine SL MoS$_2$ (Figure 5(a)) and the valence band maximum (VBM) shifts from the K point to the Γ point, which leads to an indirect gap (blue arrow, Figure 5(b)). Figure 5(c), corresponding to the case where three S atoms are replaced by three O atoms, also shows an indirect bandgap formation (blue arrow) during the oxygen plasma treatment process. Finally, the calculated bandstructure of MoO$_3$ shows that it is a wide bandgap semiconductor (see Figure 5(d)) with an indirect bandgap. Note that DFT typically underestimates the bandgap, sometimes substantially. Since pristine SL MoS$_2$ has an experimental bandgap of 1.8 eV, and pristine MoO$_3$ an experimental bandgap larger than 3.2 eV,[35] it is plausible to assume that the bandgap of SL MoS$_2$ with MoO$_3$ defects increases monotonically from 1.8 eV for 0% MoO$_3$ defect concentration to up to 3.2 eV or higher for 100% MoO$_3$ defect concentration.[36, 37] Note also that Kohn-Sham eigenstates resulting from DFT calculations are typically accurate in the sense that they provide an accurate distribution of the electron density and provide accurate results about the type of bandgap.[38] Interestingly, although pristine SL MoS$_2$ has a direct bandgap (red arrow), our calculations show that all other configurations have indirect bandgaps (blue arrow), which explains the quenching of the PL after oxygen plasma exposure. Therefore, the radiative recombination must be assisted by electron-phonon scattering inside the first Brillouin zone, which leads to the substantial reduction of the PL intensity up to quenching. In addition, our DFT calculations show that the O defect sites result in a flattening of the SL. This means that the bond angle of Mo-S, when measured from the 2D plane, is reduced. Consequently, the in-plane surface of the unit cell increases in the 2D plane, whereas the size of the unit cell in c direction gets smaller. This change in unit cell leads to tensile strain in the 2D plane and compressive strain in c direction. Since the phonon energy spectrum depends on the strain, as e.g. shown for graphene and boron nitride[39], we conclude that the energy of the phonon mode $E_{2g}^1$ in the 2D plane decreases, corresponding to a red shift, while the energy of the phonon mode $A_{1g}$ in c direction increases, corresponding to a blue shift. These conclusions agree qualitatively with the red- and blue shifts observed in the Raman spectra shown in Figure 3. This significant change in band structure is also in line with the observed quenching of the PL in Figure 2.

**CONCLUSIONS**

In conclusion, we presented strong evidence of the evolution of PL of SL $MoS_2$ via time controlled oxygen plasma exposure. The experimental results obtained with exposure time-dependence study of Raman spectroscopy, XPS and band structure theoretical calculation strongly support the hypothesis of growing lattice distortion engendered by the presence of $MoO_3$ defect sites the $MoS_2$ layer created by oxygen bombardment. Early formation (below t1) of defected $MoO_3$-disordered domains in the $MoS_2$ layer was confirmed to be one of the physical behavior causing PL decrease to quenching and significant Raman band shifting. Our results strengthen the understanding of fundamental physical properties of $MoS_2$ treated under oxygen plasma and demonstrate a new technique to engineer the bandgap and control the optical properties of atomically thin SL $MoS_2$.

**SUPPORTING INFORMATION AVAILABLE:** (1) Fitting of the time dependent photoluminescence spectra of single-layer (SL) $MoS_2$. (2) X-ray photoelectron spectroscopy (XPS) spectra of pristine and 6s plasma treated sample. This information is available free of charge via Internet at http://pubs.acs.org.

**AUTHOR INFORMATION**
**Corresponding Authors.**
*Prof. Laurene Tetard, E-mail: Laurene.Tetard@ucf.edu, Tel: 407-882-0128
*Prof. Saiful I. Khondaker, E-mail: saiful@ucf.edu, Tel: 407-864-5054, Fax: 407-882-2819

**ACKLOWDEGEMENT**
S.I.K acknowledges financial support from NSF (grant 1102228). M.N.L. acknowledges support from NSF (grant ECCS-0901784), AFOSR (grant FA9550-09-1-0450), and NSF (grant ECCS-1128597).

# Supporting Information

# Photoluminescence Quenching in Single-layer MoS$_2$ via Oxygen Plasma Treatment

Narae Kang[1,2], Hari P. Paudel[1,2], Michael N. Leuenberger[1,2,3], Laurene Tetard[1,2,*], and Saiful I. Khondaker[1,2,4,*]

[1]Nanoscience Technology Center (NSTC), [2]Department of Physics, [3]College of Optics and Photonics (CREOL), [4]School of Electrical Engineering and Computer Science, University of Central Florida, 12424 Research Parkway, Suite 400, Orlando, Florida, 32826, USA

[*]Corresponding authors

Prof. Laurene Tetard, E-mail: Laurene.Tetard@ucf.edu, Phone: 407-882-0128,

Prof. Saiful I. Khondaker, E-mail: saiful@ucf.edu, Phone: 407-864-5054, Fax: 407-882-2819,

1. **Fitting of the time-dependent photoluminescense (PL) spectra of single-layer (SL) MoS$_2$.**

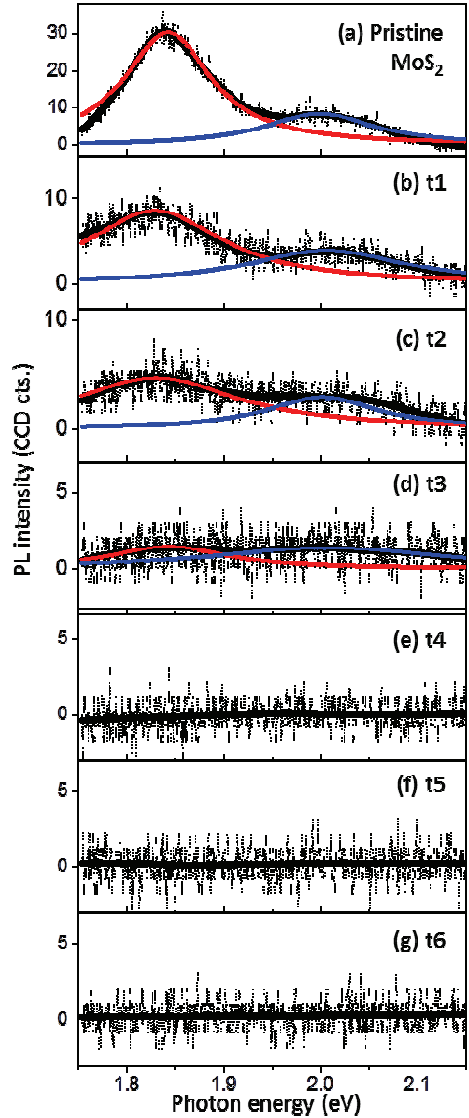

**Figure S1**. Time-dependent plasma-induced photoluminescence (PL) plots from pristine SL MoS$_2$ to t6. The original PL data is plotted as dotted line in the background as a contour plot. For each PL curves, the two prominent PL peaks, A1 peak at 1.84 eV (red curve) and B1 peak at 2.02 eV (blue curve), were fitted with a Lorentzian function in order to acquire more accurate peak intensity for both A1 and B1 peak and to compare the intensity as a function of plasma time exposure. The peaks were found and fitted until t3, however, from t4 to t6 no peaks could be detected, indicative of complete PL quenching from t4 to t6.

2. **X-ray photoelectron spectroscopy (XPS) spectra of pristine and 6s plasma treated sample.**

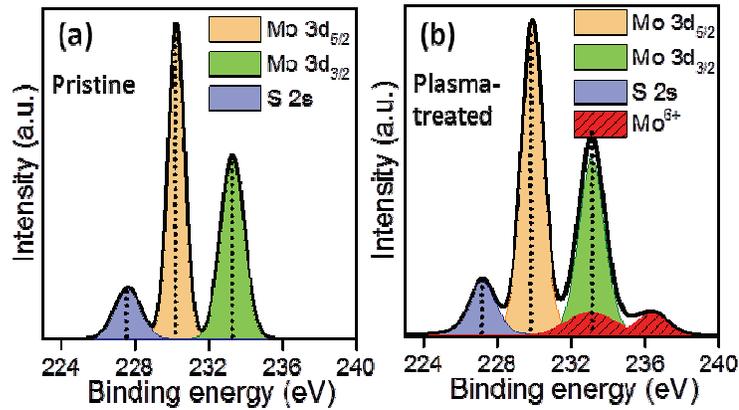

**Figure S2**. XPS spectra of Mo (3d) and S (2s) core levels for (a) before and (b) plasma treated SL MoS$_2$ flake. XPS was carried out on Physical Electronics 5400 ESCA system utilizing a monochromatized Al Kα X-ray source. The sample of MoS$_2$ that was used for XPS contains both single- and multi-layer MoS$_2$ flakes exfoliated onto SiO$_2$ substrate. The measurements were acquired before and after plasma treatment. All the peaks were fitted using Gaussian function. The convolution of original XPS data is plotted as contour curve (black), and each fitted peaks were plotted in the figure showing the area of these fitted functions measured for the integrated XPS intensity for four prominent peaks, Mo 3d$_{5/2}$ (orange), Mo3d$_{3/2}$ (green), S 2s (blue), and Mo$^{6+}$ (red). By comparing the spectra obtained before and after plasma treatment, three prominent peaks (Mo 3d$_{5/2}$, Mo3d$_{3/2}$, and S 2s) were found at 227 eV, 229.7 eV, and 233.1 eV, respectively. However, an additional peak at 236.4 eV was observed corresponding to the higher oxidation state of Mo$^{6+}$ state in the MoO$_3$ phase.[1]